\documentclass{article}%
\usepackage{amsmath}
\usepackage{amsfonts}
\usepackage{amssymb}
\usepackage{graphicx}

\begin{document}

\begin{center}{\Large \textbf{
A Universal Kinematical Group\\ for Quantum Mechanics\\
}}\end{center}

\begin{center}
Gerald A. Goldin
\end{center}

\begin{center}
Dept. of Mathematics, Dept. of Physics \& Astronomy,\\Rutgers University, New Brunswick NJ, USA

\smallskip
geraldgoldin@dimacs.rutgers.edu
\end{center}

\begin{center}
David H. Sharp
\end{center}

\begin{center}
Theoretical Division,\\
Los Alamos National Laboratory,\\
Los Alamos NM, USA

\smallskip
how.pacifica@gmail.com
\end{center}

\begin{center}
\today
\end{center}

\section*{Abstract}

In 1968, Dashen and Sharp obtained a certain singular Lie algebra of local
densities and currents from canonical commutation relations in nonrelativistic
quantum field theory. The corresponding Lie group is infinite dimensional:
the natural semidirect product of an additive group of scalar functions with
a group of diffeomorphisms. Unitary representations of this group describe a
wide variety of quantum systems, and have predicted previously unsuspected
possibilities – notably, anyons and nonabelian anyons in two space dimensions.
We present here foundational reasons why this semidirect product group serves
as a universal kinematical group for quantum mechanics. We obtain thus a
unified account of all possible quantum kinematics for systems with mass in an
arbitrary physical space, and clarify the role played by topology in quantum
mechanics. Our development does not require quantization of classical phase
space; rather, the classical limit follows from the quantum mechanics. We also
consider the relationship of our development to Heisenberg quantization.

\section{Introduction and background}
\label{sec:intro}

\subsection{Local current algebra and an infinite-dimensional group}

In developing an approach to hadron dynamics based on local current algebras, Dashen and Sharp in 1968  \cite{DasSha1968} obtained a certain singular commutator algebra for mass and momentum density operators in the physical space $\mathbb{R}^3$. They based their formal construction on canonical commutation relations in nonrelativistic quantum field theory, descriptive of bosonic particles. The same singular algebra is obtained if one begins with canonical anticommutation relations, descriptive of fermionic particles \cite{ShaWig1974}.

An initial question, then, was how systems of indistinguishable particles satisfying different exchange statistics were to be described by means of such a current algebra. The answer lay in the identification and classification of its self-adjoint representations in Hilbert space. 

Interpreting the local currents as operator-valued distributions, one obtains a bona fide infinite-dimensional Lie algebra by regularizing the singular currents. This algebra is the natural semidirect sum of a commutative algebra of real-valued scalar functions on $\mathbb{R}^3$, and an algebra of vector fields on $\mathbb{R}^3$ endowed with the Lie bracket. The associated Lie group, obtained by exponentiating the current commutators, is the natural semidirect product of an additive group $\mathcal{S}$ of scalar functions on $\mathbb{R}^3$ under pointwise addition, with a group of diffeomorphisms $\mathcal{K}$ of $\mathbb{R}^3$ under composition  \cite{GolSha1970, Gol1971}.

In fact $\mathbb{R}^3$ may be replaced by a more general physical space $M$. This may be a smooth manifold, or a manifold with boundary and/or singular points. The topology of $M$, as described by its homotopy, may also be nontrivial. Thus the group we are presently discussing is $\mathcal{S}(M)\rtimes \mathcal{K}(M)$.

Over the next fifteen years or so, it became clear that the various inequivalent irreducible unitary representations of this semidirect product group describe a wide variety of quantum systems (noted below). They also predicted previously unsuspected possibilities: anyons and nonabelian anyons in two space dimensions. This raises the question of what that diversity means, and why it should be so. One might even see it as a disadvantage of $\mathcal{S}(M)\rtimes \mathcal{K}(M)$. Indeed, the uniqueness (up to unitary equivalence) of the self-adjoint, irreducible representation of Heisenberg algebra, suggesting that Heisenberg quantization of classical mechanics provides a well-defined outcome, has been generally seen as advantageous.

The present article provides a simple and fundamental explanation for why this variety occurs. Our reasoning is quite independent of the derivation of $\mathcal{S}(M)\rtimes \mathcal{K}(M)$ from local currents obtained in turn from quantized fields and the associated formalism. We thus come to understand $\mathcal{S}(M)\rtimes \mathcal{K}(M)$ as a {\it universal kinematical group} for quantum systems with mass.

Why does this matter? We obtain quantum mechanics directly from the physical space $M$ and fundamental principles. We do not quantize a classical phase space, but we can identify such a phase space for each unitary representation of a single group. We unify the description of quantum mechanical systems of many different kinds. We suggest taking new kinematical possibilities seriously. And we clarify the kinematical role of topology in quantum mechanics, as distinguishable from its dynamical role. This perspective may be valuable to our understanding of anyonic excitations, quantum vortex configurations, and topological quantum theory generally.

\subsection{Describing a variety of quantum systems}

Let us list briefly the distinct quantum systems, some previously well-known and others that were newly predicted by this theoretical method, that are described kinematically by irreducible unitary representations of the group $\mathcal{S}(M)\rtimes K(M)$:
\begin{itemize}
\item{$N$-particle quantum mechanics, with particles distinguished by their masses \cite{Gol1971}}
\item{systems of indistinguishable particles satisfying Bose or Fermi exchange statistics in two or more space dimensions} \cite{GolMenSha1980}
\item{systems of infinitely many particles in locally finite configurations, corresponding to an infiinite (free or interacting) Bose gas \cite{GolSha1970, GolGroPowSha1974, AlbKonRoc1999}}
\item{infinitely many particles in locally finite configurations, corresponding to a Fermi gas \cite{Men1974a, Men1974b, MenSha1975}}
\item{indistinguishable particles satisfying earlier-suggested ``parastatistics" \cite{Gre1953, MesGre1964}, i.e., higher-dimensional representations of the symmetric group, under exchange in two or more space dimensions}
\item{indistinguishable particles satisfying intermediate, anyon statistics in two space dimensions \cite{LeiMyr1977, GolMenSha1980, GolMenSha1981, Wil1982a, Wil1982b}}
\item{distinguishable anyons in two space dimensions, and systems of nonabelian anyons in two space dimensions, first predicted in this way \cite{GolMenSha1985}}
\item{tightly bound composites, or particles with internal structure: point dipoles, quadrupoles, etc. \cite{GolMen1985}}
\item{particles with spin, arranged in spin towers according to representations of the general linear group \cite{GolSha1983a, GolMenSha1983}}
\item{particles with fractional spin in two space dimensions \cite{GolSha1983b}}

\end{itemize}
A recently-published historical review discusses the predicton of anyons and some of the most important ideas that were antecedent to their prediction \cite{Gol2023}. 

Over the years, numerous possibilities and applications beyond those listed above have emerged from the quantum-theoretic study of groups of diffeomorphisms and their representations. A partial list includes:

\begin{itemize}

\item{quantum systems in the presence of electromagnetic fields \cite{MenSha1977}}
\item{systems of infinitely many particles with accumulation points, and rarefied systems of infinitely many particles having zero mean density \cite{GolMos1995, GolMosSak2005}}
\item{$q$-commutation relations of quantum fields derived by interpolating hierarchies of representations of $\mathcal{S}(\mathbb{R}^2)\rtimes \mathcal{K}(\mathbb{R}^2)$ \cite{GolSha1996}}
\item{quantized vortices in two- and three-dimensions, under the constraint of incompressibility \cite{GolMenSha1987, GolMenSha1991}}
\item{generalized nonlinear quantum mechanics and associated nonlinear gauge transformations \cite{DoebGol1992, DoebGol1996, DoebGolNat1999}}
\item{generalized nonlinear, gauge-invariant time-evolutions for ``hydrodynamical'' quantum variables in the presence of electromagnetic and other external fields \cite{Gol2008}}
\end{itemize}

\noindent
In the next section, we show how the Lie algebra of local currents, and the associated group $\mathcal{S}(M)\rtimes \mathcal{K}(M)$, follow from just a few basic physical assumptions. We thus answer the question of why this method is so powerful and universal, wihout any formalism beyond the familar constructs of quantum mechanics.

In Sec. 3, we provide an overview of the representation theory. We draw attention to the space $\mathcal{S}^{\,\prime}(M)$, the dual space of $\mathcal{S}$, which contains the theoretically possible configuration spaces as subsets. We point out how the role of topology enters naturally and inevitably, and how this led to our predictions of anyon and nonabelian anyons in the early 1980s. We point out how a ``classical phase space'' may be identified for any particular irreducible unitary representation. The problem of ``quantization'' is then seen not as fundamental, but simply as the inverse problem to taking the classical limit of a quantum system.

In Sec. 4, we discuss the relationship of this picture to Heisenberg quantization. Sec. 5 is the conclusion.

\section{Foundation for a universal group predicting possible quantum kinematics}

In this section, we present ideas in a very elementary way. Our aim is to demonstrate clearly how and why simple physical assumptions imply the universality of $\mathcal{S}(M)\rtimes \mathcal{K}(M)$ in describing quantum kinematics.

\subsection{Fundamental physical asssumptions}

We begin with the following assumptions:

\smallskip
\noindent
1. We are describing an arbitrary quantum system with mass, in a physical space $M$ of dimension $n$. We take $M$ to be a smooth $(C^\infty)$ finite-dimensional Euclidean manifold, or a manifold with boundary. $M$ is equipped with a volume measure, and with the usual tools of differential geometry (e.g., a connection represented by the usual gradient symbol $\nabla$). We shall allow also for the possibility of some singular points in $M$.

\smallskip
\noindent
2. Observables in quantum mechanics are described in the usual way, by self-adjoint operators in a Hilbert space, with the usual interpretations.

\smallskip
\noindent
3. Measurements may be taken of the mass density in bounded regions of $M$. All such measurements may be taken simultaneously -- that is, the self-adjoint operators describing them commute.

\smallskip
\noindent
4. Measurements may be taken of the momentum density in bounded regions of $M$. Momentum density refers to the instantaneous transport of mass density.

\smallskip
\noindent
5. The self-adjoint operators describing momentum density measurements do not commute with those describing mass density measurements.

\smallskip
Note that in assuming the physical space $M$, we make no commitment to a spacetime symmetry group or to covariance of the currents. We call this approach ``nonrelativistic.'' That is, we allow for either Galilean quantum mechanics or for relativistic theories. Whether the spacetime symmetry is Galilean, Minkowskian, or something else will determine the possbile choices of a Hamiltonian compatible with that symmetry. In a representation by self-adjoint operators of the observables assumed above, it turns out that we are able to express Galilean or Minkowskian Hamiltonians in terms of those operators \cite{GolSha2019a,GolSha2019b}.

From the above assumptions, there follows the inevitability of quantum kinematics described by a unitary representation of the group $\mathcal{S}(M)\rtimes \mathcal{K}(M)$. Let us consider how this comes about. In the development that follows, we first show that describing measurements of the mass density in $M$ entails a unitary representation $U$ of $\mathcal{S}(M)$. We next show that describing measurements of the momentum density entails a unitary representation $V$ of $\mathcal{K}(M)$. Then the assumption that momentum refers to the instantaneous transport of mass implies that $U$ and $V$ satisfy the semidirect law for a unitary representation of $\mathcal{S}(M)\rtimes \mathcal{K}(M)$. Finally, we introduce the physical constant $\hbar$ as a physically measurable coefficient in the corresponding Lie algebra of observables. Our development makes use of some standard mathematics from the theory of distributions.

\subsection{Measurements of mass density: the commutative Lie
group $\mathcal{S}(M)$}

First, let us describe measurement of the mass density of a quantum system in
$M$ as a system of observables described by self-adjoint operators in a Hlilbert
space $H$.

Allowing for the possibiltiy of point-like particles, mass densities in $M$ may
be singular – i.e., distributions rather than functions. For example, consider the
mass density of a system of $N$ point particles, having masses $m_1, ... m_N$, located
respectively at positions $x_1, ... x_N$ in $M$. The mass density would be $\Sigma_{j=1}^N m_j \delta_{x_j}$,
where $\delta_{x_j}$ is a “Dirac $\delta$-function” centered at $x_j$ – that is, an evaluation functional. Such a density is understood as a continuous linear map from a vector space of smooth functions on $M$ (the “test functions”) to $\mathbb{R}$. Of course, mass densities may also be non-singular – in principle, a mass density may be any distribution on $M$.

Thus, by our first three physical assumptions, for a quantum system in $M$
with mass, we must represent the mass density observables in a Hilbert space
$\mathcal{H}$ by means of a self-adjoint operator-valued distribution on $M$. The function space for such a distribution will be smooth $(C^\infty)$ real-valued functions
on $M$. Restricting to ``local'' measurements, i.e. those that can be taken in compact regions of $M$, means that the test functions should have compact support – i.e., each
function is zero outside of a compact region. We denote this operator-valued
distribution describing mass density observables by $\rho(x)$; and for the test function
$f : M \to \mathbb{R}$, we write formally (in the usual notation for distributions)
$\rho(f) = \int_M \rho(x)f(x)d^nx$. Then $\rho(f)$ is the observable (self-adjoint operator)
that measures the mass density averaged over some region by the test function
$f$. Our assumption that such measurements may be made simultaneously, to
arbitrary precision, means that these operators are mutually commutative: for
any two test functions $f_1$ and $f_2, [\rho(f_1), \rho(f_2)]$ = 0.

The infinite-dimensional space of $C^\infty$ real-valued, compactly-supported functions
on $M$ is commonly denoted $\mathcal{D}(M)$ in the theory of distributions. For
mathematical reasons, when $M$ is noncompact (for example when $M = \mathbb{R}^n$) it
is sometimes helpful to work with the rather larger test-function space of $C^\infty$
real-valued functions that decrease to zero rapidly at infinity (i.e., faster than the
growth of any polynomial). This space, denoted by $\mathcal{S}(M)$, is generally known in the theory of distributions as Schwartz space. As functions in $\mathcal{S}(M)$ can be
approximated (uniformly in all derivatives) by functions in $\mathcal{D}(M)$, we may take
$\rho(f)$ to be defined as an observable for all $f \in\mathcal{S}(M)$ without thereby imposing
any restriction on the quantum theory. In principle, measurements described
by $\rho(f)$ may be approximated to arbitrary accuracy by measurements localized
in compact regions.

Finally let consider how the above applies when $M$ is a manifold with smooth
boundary $\partial M$, and/or when $M$ contains some singular points where it is not
smooth. When we want to consider a system where mass may be located on
$\partial M$, we simply take $\mathcal{S}(M)$ to include functions that are defined and are also $C^\infty$
on $\partial M$. Then $\rho(f)$ will detect a finite mass distribtution on the boundary when $f$ is nonvanishing there. In the situation that $M$ contains singular points or
“defects” (e.g., a subset of $M$ where the manifold is not differentiable), our approach
should depend on the kinematical properties associated with such points.
Deleting the subset of all such points from $M$ may change the topology of $M$ –
its homotopy – which, as we shall see below, implies new physical possibilities. This amounts to assuming that the defects are impenetrable by mass.
Alternatively, we may keep such points but require our test functions to vanish
smoothly in all derivatives as any such point is approached. This excludes the possibility
of measuring mass located at singular points. Finally, we may choose to
allow arbitrary test functions on $M$ that are not only $C^\infty$ outside the singular
subset, but are continuous and approach a well-defined limit as
such points are approached. With such test functions, one can in principle measure
the mass content of singular points in the manifold.

To sum up, the commutator algebra of operators representing mass density
observables in $M$ must be a self-adjoint representation $\rho$ of the (infinite-dimensional) function space $\mathcal{S}(M)$ endowed with a commutative bracket.

Moving from this Lie algebra to the corresponding Lie group is straightforward.
Stone’s theorem associates with each self-adjoint operator $\rho(f)$ a
strongly-continuous one-parameter unitary group $U_s(f) = \exp{is\rho(f)}, s \in \mathbb{R}$, noting here that $s\rho(f) = \rho(sf)$; and conversely, such a one-parameter group is generated by the self-adjoint operator
$\rho(f) = \mathrm{lim}_{s \to 0} (1/is)[U_s(f) - I]$.
The bracket $[\rho(f_1), \rho(f_2)] = 0$ implies that $U(f_1)U(f_2) = U(f_1 + f_2)$. Thus the mass density observables generate and are recovered from a continuous
unitary representation $U(f)$ in $\mathcal{H}$ of the abelian group $\mathcal{S}(M)$ under
pointwise addition.

\subsection{Measurements of momentum density: the diffeomorphism
group $\mathcal{K}(M)$}

Next we consider the implications of our final two assumptions, pertaining to the
measurement of momentum density. Since the mass density of the quantum system
may be singular, the momentum density may likewise be singular. Furthermore,
the momentum density is a vector quantity. Consequently, momentum
measurements must be represented in $\mathcal{H}$ as an $n$-component operator-valued distribution $J = (J_1, ...J_n)$, where $n$ is again the dimensionality of $M$.
The test functions for momentum observables are now $C^\infty$ tangent vector
fields $\mathbf{g}$ on $M$. That is, for $x \in M$, $\mathbf{g}(x)$ is an element of $T_x(M)$, the tangent
space to $M$ at $x$. The self-adjoint operator $J(\mathbf{g})$ describes the averaged momentum
density, in the direction of $\mathbf{g}(x)$ at each point $x$ and weighted there
with the magnitude $|\mathbf{g}(x)|$. We then write formally, in the usual notation for
distributions, $J(\mathbf{g}) = \int_M \mathbf{J}(x) \cdot \mathbf{g}(x)d^nx$; where $\mathbf{J}(x) \cdot \mathbf{g}(x) = \Sigma_{k=1}^n \mathbf{J}_k(x)\mathbf{g}^k(x)$.

If $M$ has a smooth boundary, we may require that for $x \in \partial M, \mathbf{g}(x)$ belongs to $T_x(\partial M)$ – that is, $\mathbf{g}$ can be nonvanishing on the boundary, but its direction there is tangential to $M$. Physically, this would mean that mass distributed on the boundary remains on the boundary; there is no component of the momentum
density in an off-boundary direction. If $M$ contains singular points (e.g., where
the tangent space is undefined), we may require $\mathbf{g}$ to vanish in all derivatives in the limit as any such point is approached. For the quantum systems we are describing, masses located at such singular points remain stationary – they belong in effect
to the ``background” in which the system is situated. Again, this implies the impenetrability of such points, and depending on their distribution in $M$, may introduce new kinematical possibiities.

When $M$ is noncompact, just as we may in principle infer the mass density
averaged with functions in Schwartz space $\mathcal{S}(M)$ from measurements averaged with compactly supported functions, we may infer the momentum density averaged with rapidly decreasing (tangent) vector fields from measurements averaged
with the vector fields having compact suport. Thus we let $\mathrm{vect}^c(M)$
denote the space of $C^\infty$ compactly supported vector fields on $M$, and $\mathrm{vect}(M)$
denote the space of $C^\infty$ vector fields on $M$ vanishing rapidly at infinity. We
then have, for $\mathbf{g} \in \mathrm{vect}(M)$, the set of self-adjoint operators $J(\mathbf{g})$ describing
measurements of momentum density.

Let us establish the commutation relations for the operators $J(\mathbf{g})$ with each
other, and identify the corresponding unitary group. These results are a consequence
of the physical interpretation of momentum as generating a flow – the
instantaneous transport of a quantity – together with the usual mathematics of
differential geometry and of self-adjoint and unitary operators.

We note that the $C^\infty$ vector field $\mathbf{g}(x)$ generates a smooth global flow on $M$.
This is the trajectory that a point would follow if it traveled along the vector
field. Such a flow is a one-parameter group of $C^\infty$ diffeomorphisms of $M$.
We denote this flow $\phi^{\,\mathbf{g}}_
r : M \to M$, where $r$ is a real parameter.
It satisfies the
differential equation $\partial \phi^{\,\mathbf{g}}_r (x)/\partial r = \mathbf{g}(\phi^{\,\mathbf{g}}_r (x)$, with the initial condition $\phi^{\,\mathbf{g}}_{r=0}(x) = x$. The flow is defined for all $x$ and for all $r$ due either to the compactness of the support of
$\mathbf{g}$, or to the controlled behavior of $\mathbf{g}$ at infinity. The product law for such a one-parameter group involves the composition of diffeomorphisms: $\phi^{\,\mathbf{g}}_{r_2} (\phi^{\,\mathbf{g}}_{r_1}(x)) = \phi^{\,\mathbf{g}}_{r_1+r_2}(x)$. To first order in $r$, the point $x \in M$ moves (in local Euclidean coordinates) to $x + r\mathbf{g}(x)$.

The one-parameter group of unitary operators in $H$ generated by $J(\mathbf{g})$ is
therefore a continuous unitary represesention of this flow, which we write as
$V(\phi^{\,\mathbf{g}}_{r})$. Composing all such flows, for $\mathbf{g} \in \mathrm{vect}(M)$, leads to a continuous unitary
representation $V$ of a group of $C^\infty$ diffeomorphisms of $M$. For $\mathbf{g} \in \mathrm{vect}^c(M)$, $\phi^{\,\mathbf{g}}_{r}$ has compact support – i.e., it is always the identity map outside the support
of $\mathbf{g}$. Any product of such diffeomorphisms likewise has compact support. Since
for noncompact $M$ the vector fields $\mathbf{g}$ are taken to vanish rapidly at infinity,
the corresponding diffeomorphisms always approach the identity rapidly at infinity.
This property, together with the topology of uniform convergence in all
derivatives, defines for us the diffeomorphsim group $\mathcal{K}$ that is represented by $V$.

We may write the multiplication
law in $\mathcal{K}$ as $\phi_1\phi_2  := \phi_2 \circ \phi_1$, where $\circ$ denote the composition of maps. We adopt this notation so that in a Hilbert space representation, $V (\phi_1\phi_2) = V (\phi_1)V (\phi_2)$ without reversing the order in which we
write the diffeomorphisms.

It is straightforward that an infinitesimal flow by $\mathbf{g}_1$, followed by $\mathbf{g}_2$, succeeded
by an infinitesimal flow back by $\mathbf{g}_1$ followed by $\mathbf{g}_2$ yields an infinitesimal flow
by the Lie bracket $[\mathbf{g}_1, \mathbf{g}_2]$ of the vector fields. In the usual notation, $[\mathbf{g}_1, \mathbf{g}_2] =
\mathbf{g}_1 \cdot \nabla \mathbf{g}_2 -\mathbf{g}_2 \cdot \nabla \mathbf{g}_1$. Thus $J$ is a self-adjoint representation in $\mathcal{H}$ of the Lie
algebra $\mathrm{vect}(M)$, equipped with the Lie bracket: i.e., $[J(\mathbf{g}_1), J(\mathbf{g}_2)]$ must be
proportional to $J([\mathbf{g}_1, \mathbf{g}_2])$.

\subsection{Momentum as the transport of mass: the semidirect
product group $\mathcal{S}(M) \rtimes \mathcal{K}(M)$}

Next we consider the commutation relations of operators describing measurements
of the mass density with those describing measurements of the momentum
density. Here we make explicit use of the physical assumption that the momentum
density refers to the infinitesimal transport of the mass density. Thus the
momentum density averaged with $\mathbf{g}$ describes the infinitesimal transport of the
mass density averaged with $f$, in the $\mathbf{g}$-direction with magnitude $|\mathbf{g}|$. This means that the commutator $[\rho(f), J(\mathbf{g})]$ must be proportional to $\rho(\mathbf{g}\cdot \nabla f)$, which is the mass density averaged with the derivative of $f$ in the $\mathbf{g}$-direction -- i.e., the Lie derivative of $f$ by $\mathbf{g}$.

At the level of the unitary groups $U(\mathcal{S})$ and  $V(\mathcal{K})$, for $f \in \mathcal{S}$ and $\phi \in \mathcal{K}$,
this gives us the semidirect product group action $V(\phi)U(f) = U(f \circ \phi)V (\phi)$.
Physically, we are here asserting that the mass density measured at the location
to which it has been carried by the momentum density takes the value measured
at its original location before being transported.

Thus we have shown how the set of possible measurements of mass and
momentum density in an arbitary quantum system in the space $M$ is described
universally by a continuous unitary representation $U(f)V(\phi)$ of the semidirect
product group $\mathcal{S}(M) \rtimes \mathcal{K}(M)$. The group law in $\mathcal{S}(M) \rtimes \mathcal{K}(M)$ is

\begin{equation}
(f_1, \phi_1)(f_2, \phi_2) = (f_1 + f_2 \circ \phi_1, \phi_1 \circ \phi_2).\label{grouplaw}
\end{equation}

\subsection{Structure constants and physical units}

We have shown that the commutation relations of $\rho$ and $J$. for $f_1, f_2, f \in \mathcal{S}(M)$ and $\mathbf{g}_1, \mathbf{g}_2, \mathbf{g} \in \mathrm{vect}(M)$, must satisfy: $[\rho(f_1), \rho(f_2) = 0; [\rho(f), J(\mathbf{g}] \sim \rho(\mathbf{g} \cdot \nabla f)$; and $[J(\mathbf{g}_1), J(\mathbf{g}_2) \sim J([\mathbf{g}_1, \mathbf{g}_2])$. But we need to establish coefficients for the two proportional relationships obtained -- i.e., structure constants for the Lie algebra. By the assumptions of quantum mechanics with which we began, these cannot all be zero. Their value depends, of course, on the physical units.

It is helpful to spell out the physical dimensions of all the constructs we have
used. Here is a short list. In our notation, $\mathrm{dim}\big[X\big]$ denotes the physical dimensions of the quantity $X$, M denotes the dimension of mass, L denotes that of length, and T
denotes that of time.

\smallskip
$\mathrm{dim}\big[M\big] = \mathrm{L}^\mathrm{n}$; i.e., n is the dimension of the physical space $M$.

\smallskip
$\mathrm{dim}\big[\rho(x)\big] = \mathrm{M/L}^\mathrm{n};\,\,\,\mathrm{dim} \big[f\big] =  1; \,\,\,\mathrm{dim}\big[\rho(f)\big] = \mathrm{M}$.

\smallskip
$\mathrm{dim}\big[\mathbf{J}(x)\big] = (\mathrm{ML/T})(1/\mathrm{L}^\mathrm{n}) = \mathrm{M/L}^{n-1}\mathrm{T};\,\,\,\mathrm{dim}\big[\mathbf{g}\big] = 1; \,\,\, \mathrm{dim}[J(\mathbf{g})] = \mathrm{ML/T}$.

\smallskip
$\mathrm{dim}\big[\mathbf{g}\cdot \nabla f\big] = 1/\mathrm{L};\,\,\,\mathrm{dim}\big[\rho(\mathbf{g}\cdot \nabla f)\big] = \mathrm{M/L}$.

\smallskip
$\mathrm{dim}\big[\,[\mathbf{g_1}, \mathbf{g}_2]\,\big] = 1/\mathrm{L};\,\,\, \mathrm{dim}\big[J([\mathbf{g_1}, \mathbf{g}_2])\big] = (\mathrm{ML/T})(1/\mathrm{L}) = \mathrm{M/T}$.

\smallskip
$\mathrm{dim}\big[\,[\rho(f), J(\mathbf{g})]\,\big] = \mathrm{M}^2\mathrm{L/T}$.

\smallskip
\noindent
Now, in the relation $[\rho(f), J(\mathbf{g})] \sim \rho(\mathbf{g} \cdot \nabla f)$, we see that the nonzero constant of proportionality must have dimensions $\mathrm{ML}^2/\mathrm{T}$ in order for the dimensions of both expressions to be the same. We therefore introduce the coefficient $\hbar$ (whose magnitude is to be determined experimentally), with

\smallskip
$\mathrm{dim}\big[\hbar\big] = \mathrm{ML}^2/\mathrm{T}$.

\smallskip
\noindent
Finally, in the relation $[J(\mathbf{g}_1), J(\mathbf{g}_2)] \sim J([\mathbf{g}_1, \mathbf{g}_2])$, the nonzero constant of proportionality must also have dimensions $\mathrm{ML}^2/\mathrm{T}$ for the dimensions of both expressions to agree. We thus obtain the ``Lie algebra of local currents'' whose self-adjoint representations describe all possible quantum systems with mass in the space $M$. 

For $f_1, f_2, f \in \mathcal{S}(M)$ and $\mathbf{g}_1, \mathbf{g}_2, \mathbf{g} \in \mathrm{vect}(M)$,
\begin{eqnarray}
[\rho(f_1), \rho(f_2)] = 0,\quad\quad\,\, \nonumber\\
\,[\rho(f), J(\mathbf{g})] = i\hbar\rho(\mathbf{g}\cdot \nabla f),\,\,\, \nonumber\\
\,\quad[J(\mathbf{g}_1), J(\mathbf{g}_2)] = -i\hbar J([\mathbf{g}_1, \mathbf{g}_2]).
\label{rhoJalg}
\end{eqnarray}
The sign and magnitude of the coefficient in the third commutation relation is established from the first two by the Jacobi identity for Lie algebras.

Let us note that with $U(sf) = \exp{[is\rho(f)]}$, the exponent $s\rho(f)$ must be dimensionless. Thus the parameter $s$ has dimension $\mathrm{1/M}$. Similarly with $V(\phi^{\,\mathbf{g}}_r) = \exp{[i(r/\hbar)J(\mathbf{g})]}$, the  parameter $r$ must have dimension $\mathrm{L}$ in order that the exponent be dimensionless.

\subsection{Remarks}

Inequivalent irreducible continuous unitary representations $U(f)V(\phi)$ of Eqs.(\ref{grouplaw}) now describe the kinematics of distinct quantum systems. The positivity of mass restricts us to representations in which the spectrum of $\rho(f)$ is positive for positive test functions. Nowhere in the development do we introduce the classical phase space of a system to be quantized. Rather, we obtain quantum theories directly for systems in $M$ from the unitary representations of $\mathcal{S}(M) \rtimes \mathcal{K}(M)$. In an important sense, then, we unify description of the kinematics of many diverse systems. Of course, the dynamics -- e.g., in interacting quantum field theories -- may connect these systems.

Reasoning very similar to the above may be carried out to describe meaurements of electric charge density and electric current density. Charge density of course need not be positive, bringing additional unitary representations into consideration. The electromagnetic field to which charges and currents couple can then also contribute to momentum density measurements.

The development here moves away from the more abstract field-theoretic derivation of the
current algebra of Eqs.(2), toward a much more elementary (and accessible) foundation for it. We also see in retrospect that the term “nonrelativisic” in the title of Ref.\cite{Gol1971} should not be understood to refer exclusively to Galilean quantum mechanics, but to the absence of commitment to a spacetime symmetry group. We think this understanding may also have relevance to efforts to characterize quantum gravity group-theoretically \cite{IshKuc1985a,IshKuc1985b}.

\section{Unitary representations of the infinite-dimensional semidirect product group}

In the next subsection we briefly summarize previously-obtained results to describe some of the insights that follow from the representation theory of $\mathcal{S}(M) \rtimes \mathcal{K}(M)$. We outline the main mathematical ideas, and point to the physical meaning of important concepts. In the subsection that follows, we show how these ideas led to an early prediction of intermediate (``anyon'') exchange statistics, and to the first prediction of nonabelian anyons.

\subsection{General description of unitary representations}


In earlier work, making use of the mathematical development by Gelfand and Vilenkin \cite{GelVil1964} and generalizing Mackey's theory of induced representations \cite{Mac1968}, we have shown that under very general conditions, a unitary representation $U(f)V(\phi)$ of $\mathcal{S}(M) \rtimes \mathcal{K}(M)$
in a Hilbert space $\mathcal{H}$ may be written:
\begin{eqnarray}
[U(f)\Psi](\gamma) = e^{i<\gamma,f>}\Psi(\gamma),\quad\quad\,\, \nonumber\\
\,[V(\phi)\Psi](\gamma) \, = \, \chi_{\phi}(\gamma)\Psi(\phi \gamma)
\sqrt {\frac{{d\mu_{\phi}} }{ {d\mu}}(\gamma)}\,. \label{repUV}
\end{eqnarray}
Here the variable $\gamma$ belongs to $\mathcal{S}^{\,\prime}(M)$, the continuous dual space of $\mathcal{S}(M)$. This is the space of distributions modeled on test functions in $\mathcal{S}(M)$. It is the set of all possible mass configurations in the physical space $M$. For $\gamma \in \mathcal{S}^{\,\prime}(M)$, $<\gamma,f>$ denotes the evaluation of $\gamma$ with the test function $f$. Thus $\mathcal{S}^{\,\prime}(M)$ serves here as a universal configuration space.

There is a natural group action of $\mathcal{K(M)}$ on $\mathcal{S}^{\,\prime}(M)$, which is the dual action to its action on $\mathcal{S}(M)$ given by the semidirect product. That is, with $\phi \in \mathcal{K}(M)$, $<\phi \gamma,f> \,=\, <\gamma,f\circ\phi>$ for $f \in \mathcal{S}^{\,\prime}(M)$.

For example, we may consider $\gamma = \Sigma_{j=1}^N m_j\delta_{x_j}$ to be a sum of evaluation functionals, with $x_j \neq x_k$ and $m_j \neq m_k$ for $j \neq k$. Such distributions describe an $N$-particle representation in $M$, with the particles distinguished by their masses. Then $\phi \gamma = \Sigma_{j=1}^N m_j\delta_{\phi(x_j)}$; i.e., $\phi$ transforms the locations of all the particles in $M$. If we begin with particles having non-coincident locations in, let us say, $\mathbb{R}^3$, we can reach via diffeomorphism any other non-coincident locations in $\mathbb{R}^3$. The {\it orbit} of $\gamma$ under the action of $\mathcal{K}$ corresponds in this example to the manifold $(\mathbb{R}^3)^{\times N} - D$, where $D$ is the ``diagonal'' set in which $x_j = x_k$ for some $j \neq k$. In $\mathcal{S}^{\,\prime}(\mathbb{R}^3)$, we label this family of distributions $\Gamma^{(N)}(\mathbb{R}^3)$.
 
In Eqs.(\ref{repUV}) $\mu$ is a countably additive measure on $\mathcal{S}^{\,\prime}(M)$ which is quasi-invariant under the action of $\mathcal{K}(M)$. This important condition means that the class of measure zero sets is preserved under transformation $\mu \to \mu_{\phi}$ by any diffeomorphism $\phi$ -- essentially, the action of $\phi$ to transform $\mu$ is nonsingular. It ensures that the Radon-Nikodym derivative $[d\mu_{\phi} / d\mu] (\gamma)$ exists almost everywhere -- i.e., outside of $\mu$-measure zero sets. For example, in the $N$-particle case,  where $\mu$ is concentrated on functionals of the form $\gamma = \Sigma_{j=1}^N m_j\delta_{x_j}$, $d\mu$ is locally equivalent to the Lebesgue measure $dx_1 ... dx_N$.

The function $\Psi$ on $\mathcal{S}^{\,\prime}(M)$ takes values in an inner product space $\mathcal{W}$, which is in general a complex Hilbert space. Usually $\Psi$ is just complex-valued, so that $\mathcal{W} = \mathbb{C}$; but more generally it may have multiple components. Letting $\langle \cdot, \cdot \rangle_{\mathcal{W}}$ denote the inner product in $\mathcal{W}$, the Hilbert space $\mathcal{H}$ for the representation in Eq.(\ref{repUV}) is the space of $\mu$-square-integrable scalar- or vector-valued functions $\Psi(\gamma)$  on $\mathcal{S}^{\,\prime}(M)$, taking values in the complex numbers $\mathbb{C}$ or in a higher-dimensional inner product space: i.e., $\int_{\mathcal{S}^{\,\prime}(M)} \langle \Psi(\gamma),\Psi (\gamma)\rangle_{\mathcal{W}}d\mu(\gamma) < \infty$. The inner product in $\mathcal{H}$ is thus given by
\begin{equation}
(\Phi,\Psi) = \int_{\mathcal{S}^{\,\prime}(M)} \langle\Phi(\gamma),
\Psi(\gamma)\rangle_{\mathcal W}\,d\mu(\gamma)\,.
\label{innerproduct}
\end{equation}

Finally, $\chi$ is a {\it measurable unitary 1-cocycle} that acts on the value space $\mathcal{W}$ of $\Psi$. It is defined and satisfies the  cocycle equation
\begin{equation}
\chi_{\phi_{1}\phi_{2}}(\gamma)\,=\,\chi_{\phi_{1}}(\gamma)
\chi_{\phi_{2}}(\phi_{1} \gamma)\label{cocycle}
\end{equation}
outside sets of $\mu$-measure zero in $\mathcal{S}^{\,\prime}(M)$.

For the representation in Eqs.(\ref{repUV}) to be irreducible, $\mu$ must be what is termed an {\it ergodic} measure. This means that for any measurable set $\Gamma \subseteq \mathcal{S}^{\,\prime}(M)$ that is invariant under the action of $\mathcal{K}(M)$, either $\mu(\Gamma) = 0$ or $\mu(\mathcal{S}^{\,\prime}(M) - \Gamma) = 0$. There are two different ways in which this can occur. One possibility is that the set $\Gamma$ on which $\mu$ is concentrated is a single orbit under the action of $\mathcal{K}(M)$ -- that is, $\mathcal{K}(M)$ acts transitively on $\Gamma$. This is generally the case when the representation in Eq.(\ref{repUV}) describes a quantum system of finitely many particles.  Alternatively, it may be that an ergodic measure is concentrated on an uncountable family of $\mathcal{K}$-orbits in  $\mathcal{S}^{\,\prime}(M)$. Such measures generally provide representations  describing quantum systems having infinitely many degrees of freedom.

The invariant set $\Gamma$ on which $\mu$ is concentrated is the {\it configuration space} of the quantum system described. {\it Thus the possible configuration spaces of quantum systems emerge from the classification of irreducible group representations.}

Next let us focus on the cocycle $\chi$ in Eqs.(\ref{repUV}) and (\ref{cocycle}). Define the {\it stability subgroup} $K_\gamma \subset \mathcal{K}(M)$ ($\gamma$ fixed) to consist of those diffeomorphisms $\phi \in \mathcal{K}(M)$ for which $\phi\gamma = \gamma$. Note that with $\phi_1, \phi_2 \in K_\gamma$, Eq.(\ref{cocycle}) is a unitary representation of $K_\gamma$ in $\mathcal{W}$. When $\mathcal{W} = \mathbb{C}$, it is a {\it character} of $K_\gamma$ -- a representation by complex numbers of modulus $1$.

When we know the configuration space $\Gamma$ and the action of $\mathcal{K}(M)$ on $\Gamma$, one way to obtain a class of cocycles, when certain conditions are satisfied, is by {\it inducing} from unitary representations of $K_\gamma$.

This ``method of induced representations,'' generalizing George Mackey's theory for finite-dimensional Lie groups, realizes the Hilbert space via {\it equivariant} wave functions $\tilde{\Psi}$ on a covering space (or more generally, a fiber bundle) $\tilde{\Gamma}$ over $\Gamma$. The action of $\mathcal{K}(M)$ on $\Gamma$ lifts naturally to this covering space. The equivariance of the wave function on $\tilde{\Gamma}$ is with respect to the representation of the stability subgroup $K_\gamma$. The natural representation of $\mathcal{K}(M)$ in the Hilbert space of square integrable, equivariant wave functions on $\tilde{\Gamma}$ is then unitarily equivalent to the representation of $\mathcal{K}(M)$ in Eq.(\ref{repUV}), with a cocycle associated to the representation of $K_\gamma$. 

Very importantly, there is a natural homomorphism from $K_\gamma$ onto the fundamental group (first homotopy group) of $\Gamma$, denoted $\pi_1(\Gamma)$. This means that unitary representations of $\pi_1(\Gamma)$ induce unitary representations of $\mathcal{K}$, and in turn provide us with the self-adjoint operators for measuring momentum densities of the quantum system with configuration space $\Gamma$. Unitarily inequaivalent representations of $\pi_1(\Gamma)$ lead to inequivalent (non-cohomologous) cocycles, and in turn to inequivalent unitary representations of $\mathcal{S}(M) \rtimes \mathcal{K}(M)$.

{\it In short, different possible quantum kinematics follow as a consequence of the nontrivial topology of} $\Gamma$.

Given $\Gamma$ and the quasiinvariant measure $\mu$ concentrated on $\Gamma$, one may always choose ${\mathcal W} = \mathbb{C}$ and $\chi_\phi(\gamma) \equiv 1$ to obtain a unitary group representation on complex-valued wave functions with a trivial cocycle. This cocycle is induced by the trivial character of $K_\gamma$. For systems of indistinguishable particles in $\mathbb{R}^n$, the topology of $\Gamma$ is nontrivial. Then the fundamental group has other one-dimensional and higher-dimensional representations. This is the fundamental, kinematical origin of all the different possible exchange statistics -- e.g., bosons, fermions, and paraparticles; and when $n=2$, anyons and nonabelian anyons. We describe this briefly in the next subsection. In the induced representation picture, the equivariant wave functions take values in the universal covering space of the configuration space, and the statistics is encoded in the equivariance. In the equavalent picture of Eq.(\ref{repUV}), the wave functions take values directly on the configuration space, and the statistics is encoded in how the self-adjoint operators act.

When $n = 1$ (i.e., particles on the real line), diffeomorphisms in $\mathcal{K}(\mathbb{R})$ cannot implement an exchange of points. Indeed, the fundamental group of the $N$-particle configuration space in $\mathbb{R}$ is trivial. We see that the origin of ``exchange statistics'' in one dimension is dynamical (i.e., a consequence of different possible self-adjoint Hamiltonian operators) rather than kinematical. We also note that when the physical space $M$ itself has a nontrivial fundamental group, even the $1$-particle configuration space admits inequivalent cocycles -- and the consequent possibility of topological effects. This is the case, for example, for the well-known Aharonov-Bohm effect \cite{AhaBoh1959,GolMenSha1981}.

As a closing remark, the system of Radon-Nikodym derivatives $[d\mu_\phi/d\mu](\gamma)$ is also a real $1$-cocycle; consequently, so is its square root. Under certain conditions, such a real cocycle can characterize a class of quasi-invariant measures on the configuration space \cite{KunSil2004}. Its logarithm can also enter as the phase of a unitary cocycle. Such cocycles are not of topological origin.

In this section, we have outlined the main ideas behind the representation theory of $\mathcal{S}(M) \rtimes \mathcal{K}(M)$, with attention to the physical meaning of each concept. See also the 1975 survey by Vershik, Gelfand and Graev \cite{VerGelGra1975} for further mathematical development, especially with regard to infinite-particle configuration spaces. We have reviewed here how the class of possible quantum configuration spaces results from the universal kinematical group $\mathcal{S}(M) \rtimes \mathcal{K}(M)$.

We have seen how the topology of configuration space results in nontrivial exchange statistics and other effects. The theory of induced representations illustrates how ``multivalued'' wave functions are actually single-valued, equivariant wave functions on the universal covering space of the configuration space. If one prefers to avoid multivalued wave functions, one can equivalently work with single-valued wave functions on configuration space, describing the exchange statistics and other topological effects by the way in which the current algebra is represented. And we are able to distinguish clearly the possibilities that are of kinematical origin from those that are dynamical.

\subsection{The prediction of anyons and nonabelian anyons in two-space}

One of the most important predictions that followed from the representation theory of $\mathcal{S}(M)\rtimes\mathcal{K}(M)$ was that of anyons in 1980-81 \cite{GolMenSha1980,GolMenSha1981} and nonabelian anyons in 1985 \cite{GolMenSha1985}. In this subsection we describe that prediction concretely.

“Anyons” are quantum particles or excitations in two-dimensional space, obeying exchange statistics intermediate between bosons and fermions. They may occur as surface phenomena in the presence of magnetic flux. Their prediction has contributed to our understanding of the quantum Hall effect, and to the theory of quantum vortices. Leinaas and Myrheim first predicted such intermediate statistics in 1977 \cite{LeiMyr1977}, based on Schr{\"o}dinger quantization; our independent prediction came from the representation theory described here. Wilczek rediscovered the possibility in 1982 \cite{Wil1982a,Wil1982b}, but omitted citation of our prior work for many years; he coined the name ``anyons.''

Nonabelian anyons obey exchange statistics satisfying a noncommutative representation of the braid group, and are likewise of theoretical importance in quantum Hall phenomena. Both abelian and nonabelian anyons have found application in the development of quantum computing. In 2020 experimentalists succeeded in creating anyonic quantum excitations \cite{Bar2020,Nak2020}; in 2023, in a project sponsored by Google, researchers implemented the structure of non-Abelian braid statistics \cite{Goo2023}. The more recent articles likewise omit citation of our early discoveries. A more detailed, accurate history is described in Ref.\cite{Gol2023}. The absence of proper citation has unfortunatey left many researchers unaware of the dramatic predictive success of the group-theoretical approach described here.

Let us describe succinctly how these predictions follow from the representation theory of the preceding subsection. Set $M = \mathbb{R}^2$ as the physical space. The $N$-point configurations $\gamma = \delta_{x_1} + \dots + \delta_{x_N}$, with $x_j \neq x_k \in \mathbb{R}^2$ for $j \neq k$, constitute an orbit in $\mathcal{S}^{\,\prime}$ under the action of $\mathcal{K}$. One may think of each configuration $\gamma$ as an $N$-point subset of $\mathbb{R}^2$, describing the locations of $N$ indistinguishable particles of unit mass. A diffeomorphism $\phi \in \mathcal{K}$ simply moves the points in $\gamma$, with $x_j \to \phi(x_j)$.

If $\phi$ leaves every point in $\gamma$ fixed, or implements a permutation of the $N$ points, then $\phi\gamma = \gamma$, and $\phi$ belongs to the stability subgroup $K_\gamma$. Conversely, if $\phi\gamma = \gamma$, it must either leave every point fixed or permute the points. So it is easy to see that there is a natural homomorphism from $K_\gamma$ onto $S_N$, the symmetric group of all such permutations. This is also the case when $M = \mathbb{R}^n, n > 2$. But because $\phi$ is a diffeomorphism that becomes trivial at infinity, its action on $x_j$ encodes more information than the final point $\phi(x_j)$. It maps any reference path from $\infty$ to $x_j$ to a path from $\infty$ to $\phi(x_j)$. This lets one keep track of the homotopy class of paths from $x_j$ to $\phi(x_j)$ associated with $\phi$. In particular, for $\phi \in K_\gamma(\mathbb{R}^2)$, it encodes how the $N$ points may wind around each other as $\phi$ maps $\gamma$ to $\phi\gamma$. The group of all such windings is the well-known braid group $B_N$. In short, there is a natural homomorphism from $K_\gamma(\mathbb{R}^2)$ to $B_N$.

The above is a special case of a very general property. The fundamental group $\pi_1(\Gamma)$ based at $\gamma \in \Gamma$ is the group of equivalence classes of continuous loops based at $\gamma$, under smooth deformation. It describes a feature of the topology of $\Gamma$; namely, the presence of ``holes.'' What we have is a natural homomorphism from $K_\gamma$ onto the fundamental group. In the case of $M = \mathbb{R}^n$, the fundamental group of the configuration space for $N$ indistinguishable particles is  $S_N$ when $n \geq 3$; but when $n = 2$, it is $B_N$.

All this means that in two-space, a unitary representation of the discrete group $B_N$ immediately provides us with a continuous unitary representation of $K_\gamma$. Then we can make use of the inducing construction in the previous subsection to obtain a continuous unitary representation of the group $\mathcal{S}(\mathbb{R}^2) \rtimes \mathcal{K}(\mathbb{R}^2)$. We then obtain the families of self-adjoint operators $\rho(f)$ and $J(\mathbf{g})$ as infinitesimal generators of unitary one-parameter subgroups. These describe the kinematics of the quantum theory.

If we restrict ourselves to complex-valued wave functions, we need only consider the one-dimensional representations of $B_N$. These may be described by ordering the points in $\gamma$ lexicographically according their locations, and assigning the phase angle $\theta$ to a single counterclockwise exchange of two adjacent points. Then $e^{i\theta}$ is the ``anyonic'' phase associated with such an exchange. When $\theta = 0$ (or $2\pi$) we have bosons; when $\theta = \pi$ we have fermions; and otherwise we have the intermediate statistics of anyons. But of course higher-dimensional, nonabelian unitary representations of $B_N$ also exist, acting on multicomponent wave functions. These likewise induce unitary representations of $\mathcal{S}(\mathbb{R}^2) \rtimes \mathcal{K}(\mathbb{R}^2)$, providing the exchange statistics of nonabelian anyons.

The universality of $\mathcal{S}(M) \rtimes \mathcal{K}(M)$ as a kinematical group for quantum theory suggests that these do not exhaust the possibilities for exotic statistics. The configuration space for $N$ {\it distinguishable} particles in $\mathbb{R}^2$ is not simply connected; its fundamental group is the group of colored braids -- the subgroup of $B_N$ for which every braid returns the points to their original positions. Unitary representations of $\mathcal{S}({R}^2) \rtimes \mathcal{K}({R}^2)$ can now associate nontrivial phases (or higher-dimensional unitary operators) with loops whereby particles fully circle each other -- not ``exchange'' statistics, but a phenomenon inherent in topological quantum theory. Likewise when $M$ itself is multiply connected, the topology of $N$-particle configuration space allows induced representations of $\mathcal{S}(M) \rtimes \mathcal{K}(M)$ to describe nontrivial phases or unitary operators associated with paths circling the excluded regions, or ``holes,'' in $M$ -- as in the Aharonov-Bohm effect.

It appears that every other development of the possibility of anyon statistics has been based ultimately on an assumed dynamics -- e.g., quantization of a system governed by a Hamiltonian, or nontrivial topology in the space of Feynman paths over which one integrates to establish the system's time-evolution. The predictions that resulted from studying local current algebra representations, obtained from the unitary representations of $\mathcal{S}(M) \rtimes \mathcal{K}(M)$, depend on kinematics only, and do not depend on quantizing a classical dynamical system.

This seems to be the natural way to arrive at exotic statistics from kinematics alone. In the next section, we explore the relationship of this group-theoretical method to conventional Heisenberg quantization. 

\section{Classical limits and their quantization}

Our perspective here does not require any quantization of the coordinates describing a classical phase space. However, if we choose a continuous irreducible unitary representation $U(f)V(\phi)$ of $\mathcal{S}(M) \rtimes \mathcal{K}(M)$, we immediately have the configuration space $\Gamma$ that describes it. We may then construct a classical phase space as the cotangent bundle $T^*(\Gamma)$. A coordinatization of $T^*(\Gamma)$ allows us to see the coordinates of $\gamma \in \Gamma$  as the positional variables, and the coordinates of cotangent vectors at $\gamma$ as the momentum variables in the phase space. We can now understand the question of quantization as an ``inverse problem'' -- how can one ``quantize'' the classical phase space $T^*(\Gamma)$ to obtain the quantum mechanics described by the representation we chose.

In the example of $N$ particles in $\mathbb{R}^3$ distinguished by their masses, the configuration space is $\Gamma^{(N)}(\mathbb{R}^3)$. The classical particle position coordinates are then the $N$ points $x_j \in \mathbb{R}^3$ entering the distribution $\gamma = \Sigma_{j=1}^N m_j \delta_{x_j} \in \Gamma^{(N)}(\mathbb{R}^3)$. The classical momentum coordinates are the $N$-tuple $(p_1, ... p_N)$ labeling a cotangent vector to $\gamma$; $p_j$ is now the $3$-dimensional momentum coordinate of the particle at $x_j$. The classical phase space is now described by coordinates $(x_1,p_1; ...; x_N,p_N) \in (\mathbb{R}^3)^{\times 2N}, \,\mathrm{with}\, x_j \neq x_k \, \mathrm{for} \, j \neq k$.

In the language of geometric quantization, the group representation has provided a polarization of the classical phase space, distinguishing the positional coordinates from the momentum coordinates. Note also that the phase space thus obtained is not precisely the same as the usual classical $N$-particle phase space; in the latter, one does not typically exclude the ``diagonal'' set where $x_j = x_k$ for some $j \neq k$.

Generally speaking, the issue of quantization is how to represent some set of functions of the classical coordinates $(x_1,p_1; ...; x_N,p_N)$ by self-adjoint operators in a Hilbert space. If $f, g$ are two such functions, with $h = \{f, g\}$ their Possion bracket, let $\hat{\mathfrak{f}}, \hat{\mathfrak{g}}$, and $\hat{\mathfrak{h}}$ be the corresponding operators. One then wants the commutator bracket of $\hat{\mathfrak{f}}$ with  $\hat{\mathfrak{g}}$ to represent $\hat{\mathfrak{h}}$. In the current example, with $x_j = (x_j^1, x_j^2, x_j^3)$ and  $p_j = (p_j^1, p_j^2, p_j^3)$,
\begin{equation}
[\hat{\mathfrak{f}},\hat{\mathfrak{g}}] = i\hbar\,\hat{\mathfrak{h}},\, \,\mathrm{where} \nonumber\\
\end{equation}
\begin{equation}
\label{poisson}
h = \sum_{j=1}^{N}\sum_{\ell=1}^3 \biggl\{ \bigl(\frac{\partial f}{\partial x_j^\ell}  \bigr)\bigl(\frac{\partial g}{\partial p_j^\ell}  \bigr) - \bigl(\frac{\partial f}{\partial p_j^\ell}  \bigr)\bigl(\frac{\partial g}{\partial x_j^\ell}  \bigr) \biggr\}.
\end{equation}
When $f$ and $g$ are the coordinate functions $x_j^\ell$ and $p_k^m$ respectively, we have the Poisson bracket $\{x_j^\ell, p_k^m\} = \delta_{jk}\delta_{\ell m}$. Then the usual Heisenberg quantization, obtained by representing the positional coordinates $x_j^\ell$ by multiplication and the momentum coordinates $p_k^m$ by $-i\hbar \partial/\partial x_k^m$, satisfies Eqs.(\ref{poisson}).

The resulting quantum system is the same as that obtained by representing $\mathcal{S}(\mathbb{R}^3) \rtimes \mathcal{K}(\mathbb{R}^3)$ on the configuration space $\Gamma^{(N)}(\mathbb{R}^3)$. But the classical functions on this configuraton space that we have, in effect, ``quantized'' in the current algebra of Eqs.(\ref{rhoJalg}) are not the coordinate functions. Instead, the self-adjoint operators $\rho(f)$ represent, by multiplication operators, all the classical functions on $T^*(\Gamma)$ of the form
\begin{equation}
\label{functionsf}
f(x_1,p_1; ...; x_N,p_N) = \Sigma_{j=1}^N m_j f(x_j), \,\mathrm{for}\, f \in  \mathcal{S}(\mathbb{R}^3).
\end{equation}
And $J(\mathbf{g})$ represents all the (3-component) classical functions of the form
\begin{equation}
\label{functionsg}
\mathbf{g}(x_1,p_1; ...; x_N,p_N) = \Sigma_{j=1}^N \bigl[\Sigma_{\ell = 1}^3 g^\ell(x_j) p_j^\ell \bigr], \,\mathrm{for}\, g^\ell \in \mathcal{S}(\mathbb{R}^3),
\end{equation}
by self-adoint operators $\Sigma_{j=1}^N (\hbar/2i )\bigl[\mathbf{g}(x_j)\cdot \nabla_{x_j} + \nabla_{x_j} \cdot \mathbf{g}(x_j)\bigr]$. Eqs.(\ref{functionsf})-(\ref{functionsg}), together with Eqs.(\ref{rhoJalg}), satisfy Eqs.(\ref{poisson}); so we indeed have obtained a quantization. Since the representation of the current algebra of Eqs.(\ref{rhoJalg}) on $\Gamma^{(N)}$ is irreducible, the operators obtained through Heisenberg quantization are determined by the operators $\rho(f)$ and $J(\mathbf{g})$.

Let us consider next the case of indistinguishable particles in $\mathbb{R}^3$. One way to formulate the Heisenberg quantization is first to label the particles $1, ..., N$, and to represent the position and momentum operators as above. One imposes the condition of indistinguishability by considering only functions of the operators that are symmetric under exchange of the $N$ labels. Then the subspaces of the $N$-particle Hilbert space that carry distinct representations of the symmetric group $S_N$ are separately invariant under all such operators. In particular, one obtains representations for $N$ bosons on the subspace of wave functions totally symmetric under exchange, and for $N$ fermions on the subpace antisymmetric under odd exchanges. One can obtain parastatistics on subspaces of multicomponent wave functions carrying higher-dimensional representations of $S_N$.

In the approach based on unitary representations of $\mathcal{S}(\mathbb{R}^3) \rtimes \mathcal{K}(\mathbb{R}^3)$, one works directly from the configuration space for $N$ particles of the same mass. Configurations in $\mathcal{S}^{\,\prime}$ are of the form $\gamma = \Sigma_{j=1}^N m \delta_{x_j}$ with $x_j \neq x_k$ for $j \neq k$. We may label this configuration space $\Gamma^{(N)}_{ident}(\mathbb{R}^3)$. It corresponds to the manifold of (unordered) $N$-point subsets of $\mathbb{R}^3$. The functions that are ``quantized'' by the current algebra are just those of Eqs.(\ref{functionsf})-(\ref{functionsg}) -- defined on {\it subsets} and with all of the masses $m_j$ in Eq.(\ref{functionsf}) having the same value $m$. These functions are already symmetric, as the configurations themselves (together with their cotangent vectors) are invariant under permutation. The distinct representations associated with particle exchange emerge from the representation theory. They result from inquivalent cocycles for the action of $\mathcal{K}$, on the same configuration space. We see directly how such representations are obtained from the fundamental group of $\Gamma^{(N)}_{ident}(\mathbb{R}^3)$, which is $S_N$.

The case $M = \mathbb{R}^2$ deserves special attention. One of the authors (GG) would like to acknowledge interesting conversations with Jan Myrheim in Trondheim, in 2019 and 2022, which inspired this section of the present article \cite{Myr2019}. Myrheim pointed out the difficulty, still unresolved, with obtaining anyon statistics through Heisenberg quantization. He and Leinaas arrived at their original ground-breaking prediction, for indistinguishabe particles in $\mathbb{R}^2$, only through a Schr{\"o}dinger quantization \cite{LeiMyr1977}.

The discussion here suggests that the difficulty rests in introducing a labeling, quantizing the coordinate functions of the phase space, and then imposing the symmetry under $S_N$ as a restriction on the observables. Instead one should introduces no artificial labels, but quantize all the symmetric classical functions $f = \Sigma_{j=1}^N m f(x_j)$ and $\mathbf{g} = \Sigma_{j=1}^N \bigl[\Sigma_{\ell = 1}^2 g^\ell(x_j) p_j^\ell \bigr], \,\mathrm{for}\, f, g^\ell \in \mathcal{S}(\mathbb{R}^2)$. In fact, it is sufficient to represent functions $f$ and $\mathbf{g}$ having compact support in $\mathbb{R}^2$. Then one obtains all of the quantum kinematics for anyons and nonabelian anyons from continuous, irreducible unitary representations of $\mathcal{S}(\mathbb{R}^2) \rtimes \mathcal{K}(\mathbb{R}^2)$ and the corresponding self-adjoint representations of the current algebra. They are induced by the unitary representations of the braid group $B_N$ -- the fundamental group of the configuration space.

We remark that in a self-adoint representation of the current algebra of Eqs.(\ref{rhoJalg}), different Hamiltonian operators can be written explicitly in terms of the local currents. These include Schr\"odinger Hamiltonians describing free or interacting particles in Galileian quantum mechanics, as well as Hamiltonians compatible with relativistic quantum field theories. Schr\"odinger quantization of specific classical dynamical systems yields quantum systems whose kinematics are described by representations of $\mathcal{S}(M) \rtimes \mathcal{K}(M)$.

\section{Conclusion}

We have presented fundamental reasons why $\mathcal{S}(M) \rtimes \mathcal{K}(M)$ serves as the universal group whose unitary representations describe the quantum kinematics of all systems with mass in the physical space $M$. This universality follows from a few basic mathematical and physical assumptions. It offers a way to arrive directly at quantum systems without the need to quantize classical physics. It also provides a deeper understanding of how topology plays the role that it does in quantum kinematics, including its governing of possibilities for the exchange statistics of quantum particles.

After many years, we finally understand the basis for the previously puzzling apparent universality of this group. It is an understanding that seems more and more obvious, almost self-evident, the more one thinks about it.

\end{document}